\documentclass[preprint,aps,prl,showpacs]{revtex4}
\usepackage{multirow}
\usepackage{graphicx}
\usepackage{dcolumn}
\usepackage{bm}

\begin{document}

\title{Pb adatom activated dissociation of O$_2$ and oxidation of Pb surfaces}% Force line breaks with \\

\author{Yu Yang,$^{1,2}$ Jia Li,$^2$ Zhirong Liu,$^3$ Gang Zhou,$^2$  Jian Wu,$^2$
Wenhui Duan,$^2$\footnote{Corresponding author. E-mail:
dwh@phys.tsinghua.edu.cn} Peng Jiang,$^4$ Jin-Feng Jia,$^2$ Qi-Kun
Xue,$^{2,4}$ Bing-Lin Gu,$^2$ and S. B. Zhang$^5$}
\affiliation{$^1$LCP, Institute of Applied Physics and
Computational Mathematics, P.O. Box 8009, Beijing 100088, People's
Republic of China} \affiliation{$^2$Department of Physics,
Tsinghua University, Beijing 100084, People's Republic of China}
\affiliation{$^3$College of Chemistry and Molecular Engineering,
Peking University, Beijing 100871, People's Republic of China}
\affiliation{$^4$Institute of Physics, Chinese Academy of
Sciences, Beijing 100080, People's Republic of China}
\affiliation{$^5$Department of Physics, Applied Physics, and
Astronomy, Rensselaer Polytechnic Institute, Troy, NY 12180, USA}
\date{\today}% It is always \today, today,

\begin{abstract}
We investigate the dissociation of O$_2$ on Pb(111) surface using
first-principles calculations. It is found that in a practical
high vacuum environment, the adsorption of molecular O$_2$ takes
place on clean Pb surfaces only at low temperatures such as 100 K,
but the O$_2$ easily desorbs at (elevated) room temperatures. It
is further found that the Pb adatoms enhance the molecular
adsorption and activate the adsorbed O$_2$ to dissociate during
subsequent room temperature annealing. Our theory explains the
observation of a two-step oxidation process on the Pb surfaces by
the unique role of Pb adatoms.

\end{abstract}


\maketitle

\clearpage

It is of fundamental importance to understand the adsorption,
dissociation, and the corresponding kinetic processes of diatomic
molecules on materials surfaces \cite{Darling1995}. As a
prototypical example, the adsorption and dissociation of O$_2$ gas
on metal surfaces have attracted considerable attention because
the resulting metal oxides have been widely used as catalysts,
sensors, dielectrics, and corrosion inhibitors \cite{Henrich}. In
this regard, lead (Pb) (111) surface is a good model system,
because atomically-flat Pb(111) terraces have been fabricated on
both metal and semiconductor substrates such as Ru (0001)
\cite{Thurmer2001} and Si(111) \cite{Upton}. Rich experimental and
theoretical literatures are now available for such surfaces
\cite{Menzel}. It was found that O$_2$ molecules adsorb on Pb(111)
at temperatures as low as 100 K. Subsequent room temperature
annealing will not result in the desorption of O$_2$ but the
oxidization of the Pb surface \cite{Ma}. In contrast, the Pb(111)
surface is remarkably resistant to oxidation at room temperature
in the experimental environment \cite{Ma,Thurmer}. From simple
energetic argument, it is very difficult to reconcile these
experimental observations, namely, how can the molecularly
adsorbed O$_2$ at low temperature does not \emph{desorb} at
(elevated) room temperature annealing, while the same O$_2$
molecules would hardly adsorb on to the same surface at the same
temperature.

In this paper, we explore the mechanisms for the puzzling
adsorption and dissociation of O$_2$ on Pb(111) surfaces, based on
first-principles total-energy calculations and thermodynamic
analysis. We show that one cannot determine the state of O$_2$ at
different temperatures solely from the calculated adsorption
energy but rather one has to rely on the Gibbs free energy that
explicitly includes the contributions from gas-phase oxygen in
terms of its atomic chemical potential. Furthermore, we show that
Pb adatoms, a common surface defect on the Pb terraces, play a key
role in retaining the O$_2$ adsorbed at low temperatures and
assisting its dissociation once the temperature is increased. Our
results are in good agreement with experiments.

Our calculations are based on the density functional theory as
implemented in the Vienna {\it Ab-initio} Simulation Package
(VASP) \cite{VASP}. The Perdew-Wang-91 (PW91)\cite{PW91}
generalized gradient approximation and the projector-augmented
wave (PAW) potential\cite{PAW} are used to describe the
exchange-correlation energy and the electron-ion interaction,
respectively. The nudged elastic band (NEB) method\cite{Mills} is
used to find the minimum energy path and the transition state for
O$_2$ dissociation. We confirm that the transition states have
only one imaginary frequency. The Pb(111) surface is modeled by a
$2\times2$ surface slab with 4 atomic layers and a vacuum layer of
15 \AA~ thickness. The integration over the Brillouin zone is
carried out by using the Monkhorst-Pack scheme\cite{Monkhorst}
with a $7\times 7\times 1$ grid. The cutoff energy for the plane
wave expansion is 400 eV. All the atoms, except for those at the
bottom layer, are relaxed until the forces are less than 0.01
eV/\AA. The calculated lattice constant for bulk Pb and the bond
length for isolated O$_2$ are 5.03 \AA~ and 1.24 \AA,
respectively, in good agreement with the experimental values of
4.95 \AA\cite{Wyckoff} and 1.21 \AA\cite{Huber}. To calculate the
adsorption energy, we use the revised Perdew-Burke-Ernzerhof
(RPBE) functional for the exchange-correlation energy at the
already optimized geometries and the self-consistent electron
densities given by PW91. This scheme is known to yield better
results when compared with experiments \cite{Hammer}.

We have previously studied\cite{Yang} in-depth the molecular O$_2$
adsorption on Pb(111), which establishes the necessary initial
state for the O$_2$ dissociation. Figure 1 shows the calculated
dissociation path where the initial state is the molecular
adsorption (MA) state and the final state is the atomic adsorption
(AA) state. The MA state has no energy barrier for adsorption
\cite{Yang}, whereas the AA state is 2.29 eV lower in energy than
the MA state. In the AA state, the two oxygen atoms reside at two
adjacent surface hollow sites. The calculated energy barrier going
from MA to AA is 0.32 eV. At the transition state (TS1), the two
oxygen atoms reside on two adjacent surface bridge sites separated
by 1.80 \AA.

\begin{figure}
\includegraphics[width=0.5\textwidth]{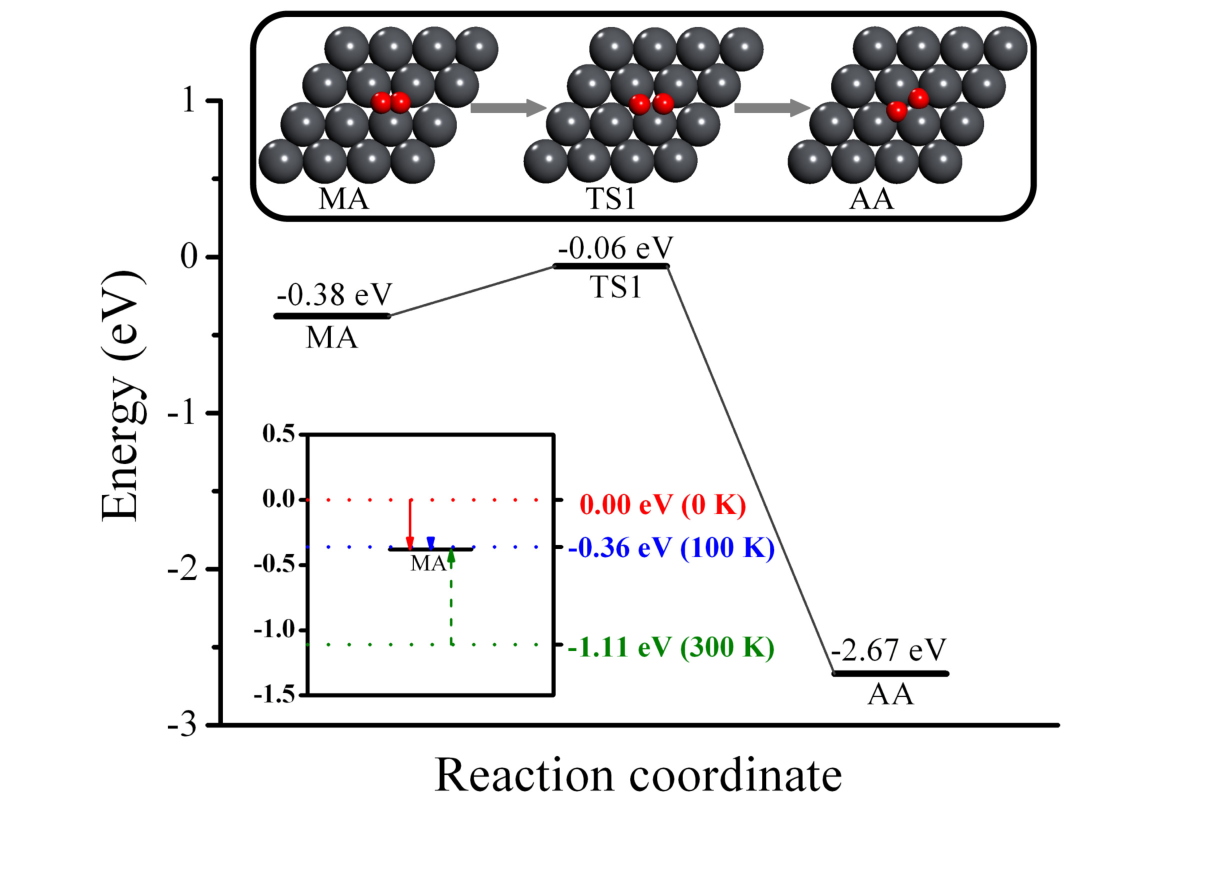}
\caption{(Color online). O$_2$ energies on clean Pb (111) surface.
Here, MA and AA stand for molecular and atomic adsorptions,
respectively, and TS1 stands for transition state between the two.
The corresponding atomic structures are given in the top panel of
the figure in which grey balls are Pb and red balls are O. The
lower inset shows the calculated Gibbs free energies for a free
O$_2$ at $T =$ 0 (red dotted line), 100 (blue dotted line), and
300 K (green dotted line), respectively. The oxygen partial
pressure in the calculation is 2 $\times$ 10$^{-7}$ Torr.}
\end{figure}

Based on the van't Hoff-Arrhenius law in the harmonic approximation
\cite{Vineyard}, the rate coefficient $k$ for the dissociation can be
expressed as \cite{Vegge}:
\begin{equation}\label{k}
k = \frac{\prod_{i=1}^{3N}\nu_{i}^{\rm
MA}}{\prod_{j=1}^{3N-1}\nu_{j}^{\rm TS1}}\times e^{-\Delta
E/k_BT}=\nu\times e^{-\Delta E/k_BT}
\end{equation}
where $\nu_{i}^{\rm MA}$ and $\nu_{j}^{\rm TS1}$ are the
frequencies of the eigenmodes at molecular adsorption and
transition states, respectively, $\nu$ is the attempt frequency
(for O$_2$, the calculated $\nu$ = 1.33$\times$10$^{13}$
s$^{-1}$), $\Delta E$ is the energy difference between the initial
and transition states, and $k_B$, $T$, and $N$ are the Boltzmann
constant, temperature, and number of atoms, respectively. Using
the barrier in Fig. 1 ($\Delta E$ = 0.32 eV), we obtain the rate
coefficients for O$_2$ dissociation: namely, $\sim$10$^7$ s$^{-1}$
at 300 K and $\sim$10$^{-4}$ s$^{-1}$ at 100 K. Hence, molecular
adsorption is stable at 100 K but the adsorbed O$_2$ easily
dissociates at 300 K. This conclusion, however, contradicts with
experiments showing that Pb(111) surfaces are resistant to
oxidation at room temperature \cite{Thurmer,Ma}.

The origin of this discrepancy is the crucial role of
thermodynamic effect in real experiments at high vacuum and
variable temperature: the Gibbs free energy $G(T,p)$, instead of
the total energy, should be used to determine the stability of the
adsorption structures. In other words, we need to include the
atomic chemical potential of oxygen, $\mu_{\rm O}$. Within the
ideal-gas law, $\mu_{\rm O}$ can be calculated
by\cite{Reuter,Zhao2005}
\begin{equation}\label{k}
\mu_{\rm O}(T,p)=\mu_{\rm O}(T,p_0)+\frac{1}{2}k_BT{\rm
ln}(p/p_0),
\end{equation}
where $p$ is the pressure and $p_0$ = 1 atm. One can find
$\mu_{\rm O}(T,p_0)$ as a function of temperature in Ref.
\onlinecite{Reuter}: at 100 K, $\mu_{\rm O}(T,p_0) = -0.08$ eV and
at 300 K, $\mu_{\rm O}(T,p_0) = -0.27$ eV. For the experimental
pressure of $p \sim10^{-7}$ Torr \cite{Thurmer,Ma}, the calculated
$\mu_{\rm O}$ at 300 K is approximately $-0.55$ eV. When taking
into account the effect of the $\mu_O$, the energy of the
gas-phase O$_2$ is $1.11 - 0.38 = 0.73$ eV lower than that of MA,
according to Fig. 1 and its inset. In other words, molecular
adsorption will not take place at room temperature. The rate
coefficient at 300 K for a gas-phase O$_2$ to dissociate with an
estimated energy barrier of 1.05 eV is quite small (about
10$^{-5}$ s$^{-1}$). So neither molecular nor atomic adsorption of
oxygen is likely to occur at this temperature. These results
explain the observations that Pb(111) surfaces resist oxidation at
room temperature \cite{Thurmer,Ma}.

From the thermodynamic viewpoint the discrepancy at room
temperature is resolved, but a new problem arises at low
temperatures. According to the free energy calculation, at
 $T =$ 100 K and $p\sim 10^{-7}$ Torr, the molecular adsorption
state of $-0.38$ eV is only 0.02 eV lower than that of a free
O$_2$ (as shown in the inset of Fig. 1). Therefore, at this
temperature the adsorbed O$_2$ will desorb easily. With $\Delta E$
= 0.32 eV in Fig. 1, the rate coefficient at 100 K of
$\sim$10$^{-4}$s$^{-1}$ is very small. Thus the adsorbed O$_2$ has
practically no chance to dissociate to stay on the surface before
it desorbs. This result, however, contradicts with the observation
that O$_2$ adsorbs on the Pb surfaces at around 100 K and oxidizes
the surface upon subsequent room temperature annealing \cite{Ma}.

To explain the experiment, we note that surface impurities can act
as nucleation centers for PbO grains from which the grains grow
rapidly and auto-catalytically \cite{Thurmer}. Pb adatoms are
commonly seen on epitaxial Pb surfaces \cite{Chang} and readily
diffuse at temperatures as low as 100 K \cite{Chan}. Therefore, Pb
adatoms could be the missing link between theory and experiment.
For the adatoms to affect the oxidation, two scenarios should be
considered: i) a diffusing Pb adatom gets to the vicinity of an
adsorbed O$_2$ and the two then interact or ii) an O$_2$ directly
adsorbs near a Pb adatom. Because the Pb adatoms are fast
diffusers that very effectively transport Pb atoms from step edges
to other parts of the terraces, we feel that the first scenario
could be more relevant to the experiment.

To study the Pb-O$_2$ interaction, we place a Pb adatom at a
distance reasonably away from the O$_2$ yet close enough for the
two to attract each other, for example, at the second nearest
neighbor bridge sites around the O$_2$. Our previous study showed
that an O$_2$ on Pb(111) has two nearly degenerate states
\cite{Yang}, and correspondingly, six different local Pb
configurations. As a result of the atomic relaxation, they all
converge to one of the two final states with shorter Pb-O$_2$
separations. Figure 2(a) shows the lower-energy configuration,
denoted as the non-dissociative adsorption (NDA) state. In the NDA
state, the O-O bond length increases from 1.24 \AA~ for isolated
O$_2$ to 1.50 \AA. This suggests that the O-O bond, although not
broken, is significantly weakened with a 21 \% elongation. The
calculated Pb-O$_2$ binding energy is 0.70 eV, indicating that the
adatoms can substantially increase the O$_2$ adsorption.

\begin{figure}
\includegraphics[width=0.48\textwidth]{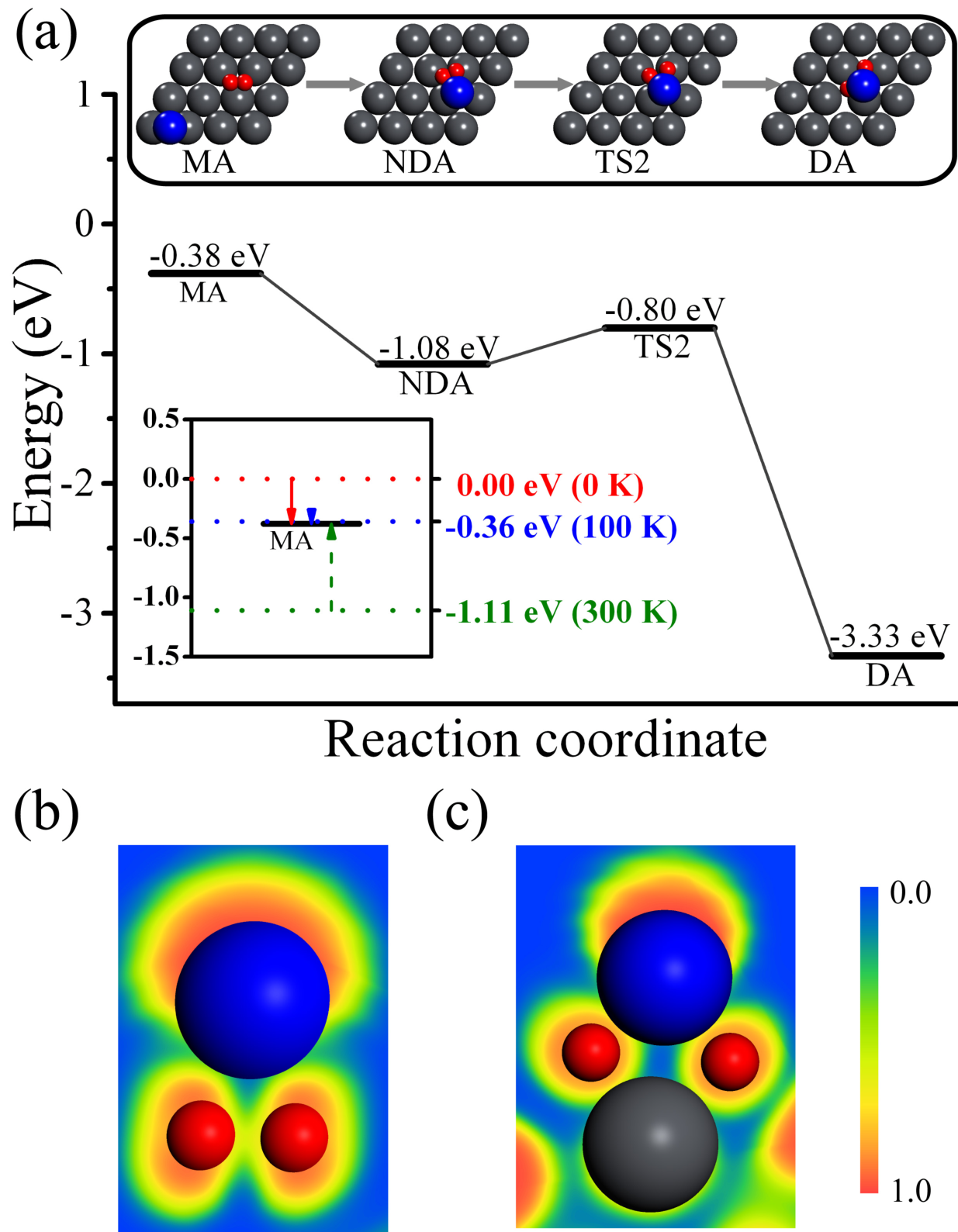}
\caption{(Color online). (a) O$_2$ energies on Pb (111) surface in
the presence of a Pb adatom. MA stands for the molecular O$_2$
adsorption far away from any Pb adatom, NDA and DA stand for
nondissociative and dissociative adsorptions, respectively, around
the Pb adatom, and TS2 is the transition state between the two.
Inset has the same meaning as the inset in Fig. 1, but for
configurations shown in Fig. 2. (b) and (c) Contour plots of the
electron localization function (see text) for NDA and DA,
respectively, in a plane containing the Pb adatom and two oxygen
atoms. Grey, blue, and red balls stand for surface Pb, Pb adatom,
and O atom, respectively.}
\end{figure}

To determine the final state of O$_2$ dissociation in the presence
of Pb adatom, we started from the NDA state and increased the
distance between the two O atoms within a plane parallel to the Pb
surface plane until there is no restoring force between them, and
then performed structural optimization. Figure 2(a) (far right)
shows the lowest-energy final state, denoted as the dissociative
adsorption (DA) state. In the DA state, the distance between two O
atoms is 2.94 \AA, which is 137\% longer than the normal O-O bond
length of 1.24 \AA. Electron localization function (ELF)\cite{ELF}
calculation confirms the dissociation. Figures 2(b) and 2(c) show
the ELFs for the NDA and DA states. One can see clearly (weak)
covalent bonding in Fig. 2(b), but not at all in Fig. 2(c). Bader
topological analysis \cite{Bader} further reveals that the two O
atoms in the DA state gain 1.99 and 1.80 electrons, respectively,
from the neighboring Pb atoms. The interactions between the O's
and Pb in the DA state are thus ionic. A slight asymmetry in the
charge transfer is because the two oxygen atoms are on
inequivalent surface sites.

Figure 2(a) shows the dissociation path for O$_2$ in the presence
of Pb adatom. It can be seen that in this case, the dissociation
barrier of 0.28 eV is much lower than the energy difference of
0.70 eV between the NDA and MA states. Note that at 300 K, the MA
state will evolve to the desorption state (discussed earlier).
Hence, an adsorbed O$_2$ is more likely to dissociate than to
desorb from the surface in this case. The calculated attempt
frequency for Fig. 2 is $\nu$ = 2.97$\times$10$^{13}$ s$^{-1}$ and
the corresponding rate coefficient at 300 K from NDA to the
transition state TS2 is $\sim$10$^8$ s$^{-1}$). Therefore,
activated by the Pb adatoms, the adsorbed O$_2$ can easily
dissociate during a room temperature annealing. These results
bring our theory into complete qualitative agreement with
experiments\cite{ Thurmer,Ma}.

In summary, we have developed a comprehensive first-principles
theory for the dissociation of O$_2$ on Pb(111) surface. It is
found that one needs to include oxygen atomic chemical potential
given by the experimental conditions (i.e., partial pressure and
ambient temperature) to qualitatively account for the observation
that O$_2$ can adsorb on clean Pb surfaces at 100 K but cannot at
room temperature. In addition, Pb adatoms, commonly seen on Pb
surfaces, can result in a substantial enhancement of the O$_2$
adsorption. As a result, O$_2$ adsorbed at low temperatures such
as at 100 K is more likely to dissociate than desorbing upon
subsequent room temperature annealing. These results explain the
recent puzzling experimental observations as well as demonstrating
the important role of surface defects in activating the oxidation
of surfaces.

This work was supported by the Ministry of Science and Technology
of China (Grant Nos. 2006CB605105 and 2006CB0L0601), and the
National Natural Science Foundation of China.

%\clearpage

\end{document}